\documentclass[twocolumn,preprintnumbers,amsmath,amssymb]{revtex4}

\usepackage{graphicx}
\usepackage{dcolumn}
\usepackage{bm}
\usepackage{color}

\begin{document}

\title{Josephson squelch filter for quantum nanocircuits}

\author{P. Forn-Diaz}
\email{p.forndiaz@tudelft.nl}
\author{R. N. Schouten}
\author{W. A. den Braver}
\author{J. E. Mooij}
\author{C. J. P. M. Harmans}

\affiliation{Quantum Transport Group, Kavli Institute of Nanoscience, Delft University of Technology, Lorentzweg 1, 2628 CJ Delft, The Netherlands}

\date{\today}

\begin{abstract}
We fabricated and tested a squelch circuit consisting of a copper powder filter with an embedded Josephson junction connected to ground. For small signals (squelch-ON), the small junction inductance attenuates strongly from DC to at least 1 GHz, while for higher frequencies dissipation in the copper powder increases the attenuation exponentially with frequency. For large signals (squelch-OFF) the circuit behaves as a regular metal powder filter. The measured ON/OFF ratio is larger than 50dB up to 50 MHz. This squelch can be applied in low temperature measurement and control circuitry for quantum nanostructures such as superconducting qubits and quantum dots.
\end{abstract}

\pacs{}

\maketitle
\begin{figure}
\includegraphics[width=\columnwidth]{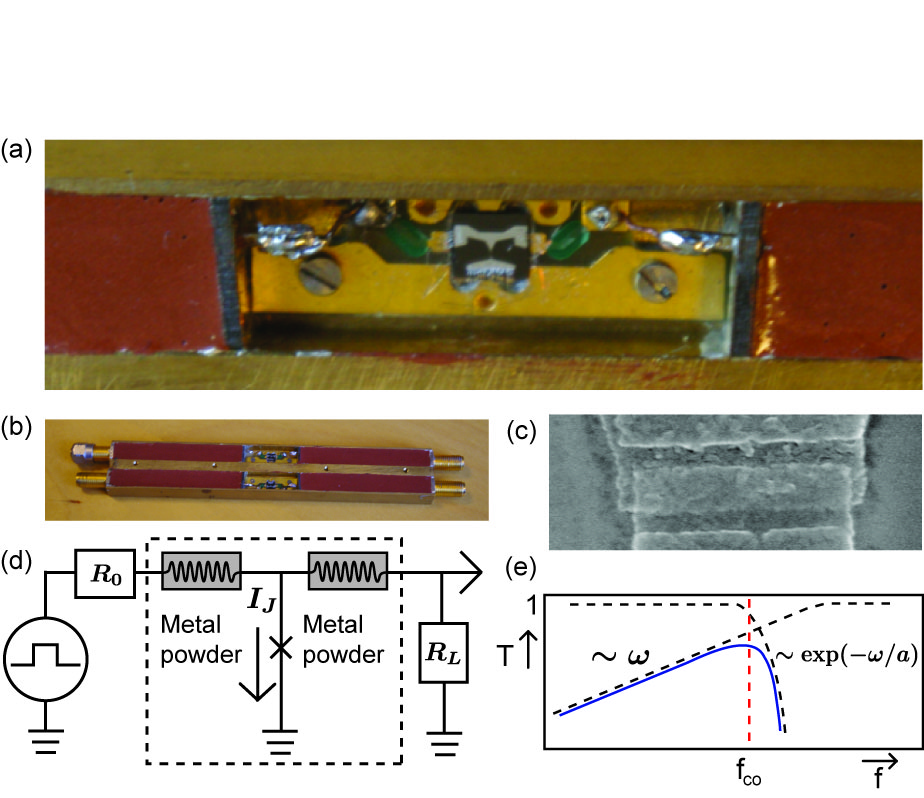}
\caption{\label{fig1}(Color online) The Josephson squelch circuit. (a) Printed circuit board with the chip containing the Josephson junction. The brownish areas to the left and right is the copper loaded epoxy containing the meandering signal wires. (b) Two complete filters with SMA connectors, built side by side. The total length is 11 cm containing 80 cm of thin Cu wire. (c) Scanning electron micrograph of a Josephson junction. The junction is fabricated using electron beam lithography and double angle shadow mask evaporation. The area is 1 $\times$ 0.2 $\mu$m$^2$ and the critical current density $j_C=15$ $\mu$A$/\mu $m$^2$. (d) Schematic of the Josephson squelch between a source of impedance $R_0$ and a load $R_L$. The Josephson junction is represented by a cross. (e) Transmission of the Josephson squelch. For low frequencies the Josephson junction acts like an inductive short, with transmission $\sim$$\,\omega$. For frequencies higher than the cutoff frequency of the copper powder filter $f>f_{co}$, the absorption of the incoming noise increases
exponentially with frequency.}
\end{figure}
\begin{figure}
\includegraphics[width=\columnwidth]{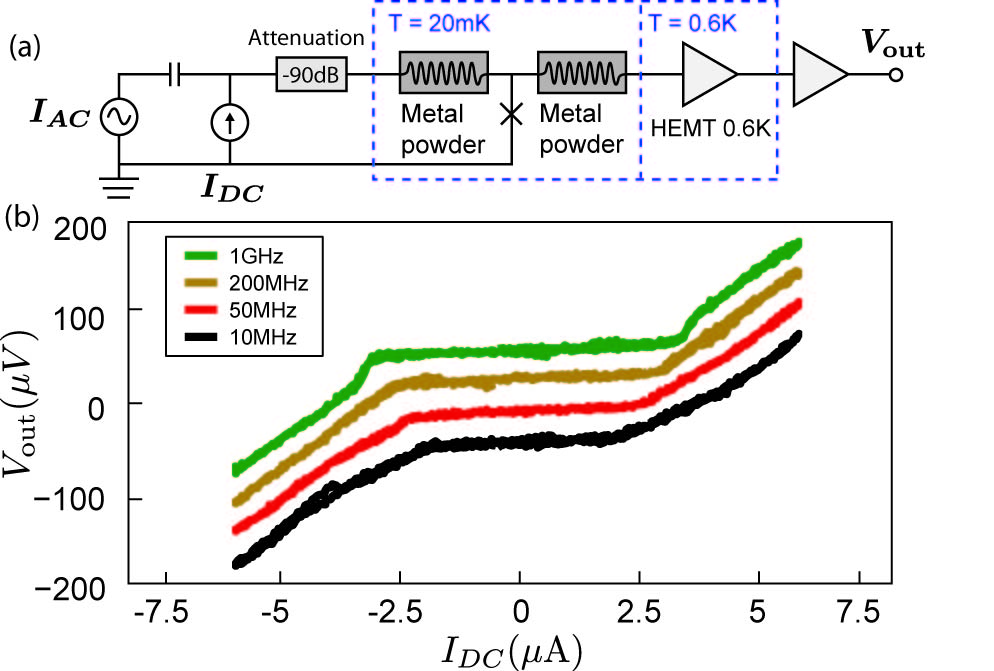}
\caption{\label{fig2}(Color online) Test circuit for the Josephson squelch and characterization of the attenuation in the copper powder filter. (a) Schematic of the electrical circuit. The response of the filter is sensed by a low temperature HEMT amplifier whose output is amplified at a room temperature post-amplifier. (b) Current-voltage characteristic. Dif\mbox{}ferent traces are vertically displaced for clarity. Each trace is taken with a 40kHz current ramp $I_{DC}$, while an AC signal $I_{AC}$ at a fixed power and different frequency is added. For frequencies below the cutoff frequency of half the copper powder filter (where the junction is located) $2f_{co}=200$ MHz the AC signal $I_{AC}$ is not attenuated, and less DC current $I_{DC}$ is needed to reach the critical current $I_C$.}
\end{figure}
\begin{figure}
\includegraphics[width=\columnwidth]{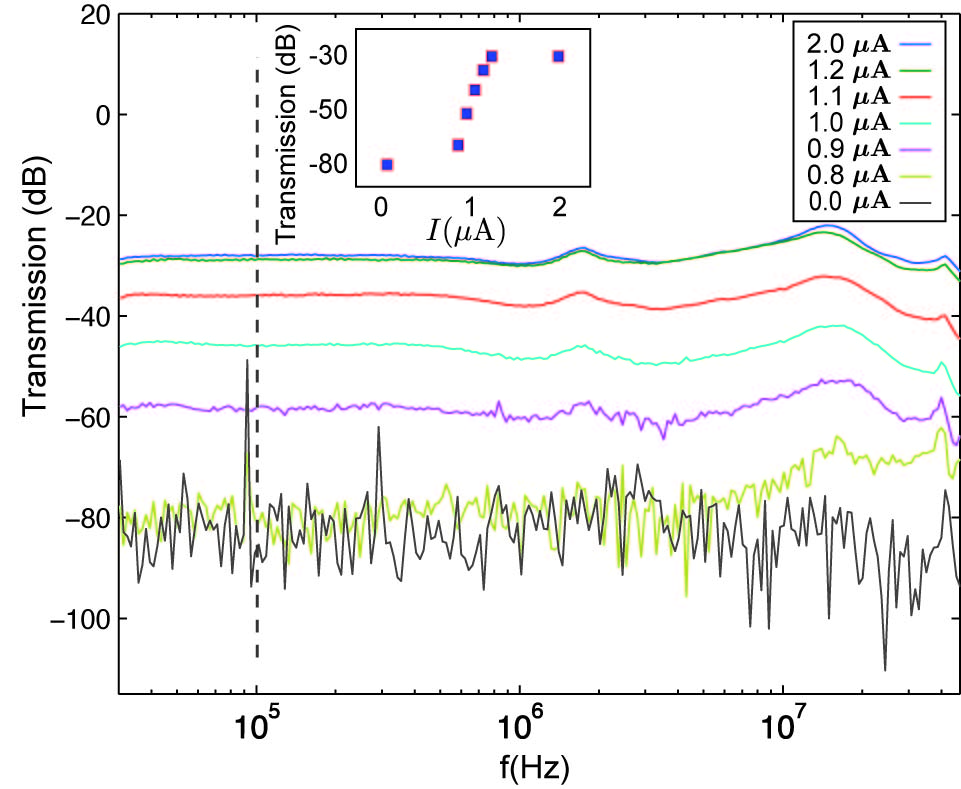}
\caption{\label{fig3}(Color online) Squelch action of a Josephson junction with $I_C\approx 1\mu$A. The transmission of the system is measured with an applied input power of $P_{\mathrm{in}}=-90$ dBm for different values of the DC current. The topmost two traces fall together because the junction is in its dissipative state. At zero-current the transmission is the noise floor of the measurement. (Inset) Cut of the amplitude at the frequency of 100 kHz (vertical dashed line). Above 1.2 $\mu$A the DC current switches the Josephson squelch OFF. The ON/OFF ratio is 50dB.}
\end{figure}

Quantum nanocircuits, such as superconducting qubits or semiconductor quantum dots, are nonlinear systems that are adversely influenced by noise. External noise sources are unavoidably introduced by connection lines for electrical or magnetic bias, operational signals and measuring devices. Commonly these lines are permanently coupled to the quantum circuit. Low temperature filtering is used, but that leads to signal loss. We have developed a squelch circuit, a nonlinear filter that shorts all low amplitude signals from DC to high frequencies but transmits with low loss when a stronger signal is applied. It consists of a Josephson junction that acts as a nondissipative short for any current below its critical current, embedded in a lossy transmission line.\\
\indent Filters based on lossy dielectrics using small metal grains have been extensively employed in the field of quantum nanocircuits \cite{martinis},\cite{vion},\cite{fukushima},\cite{bladh},\cite{rsikoch},\cite{slichter}. Such filters consist of a metallic wire embedded in a metal powder. Via induced eddy-currents in the grains, the energy of a signal propagating along the wire at higher frequencies is dissipated. Transmission starts at DC with negligible attenuation, while beyond a cut-off frequency $f_{co}$ of typically 0.1 GHz the attenuation very steeply increases in a roughly exponential way with frequency. The uniform and distributed nature of these low pass filters makes them behave smoothly versus frequency, maintaining very small transmission far beyond cut-off.
The Josephson junction \cite{orlando} is the element in our filter responsible for the squelch action. It has a current-voltage (I-V) characteristic which shows zero DC voltage for currents $|I_J|$ up to a critical current $I_C$, and acts as an inductance that for low currents is given by $L_J=\Phi_0/2\pi I_C$. For $|I_J|>I_C$ the voltage $V$ shows an approximately linear increase with current. Thus, the junction can be employed as a switch, as for $|I_J|<I_C$ it has a low impedance $2\pi fL_J$, while for $|I_J|>I_C$ the effective impedance is approximately given by the Josephson junction normal state resistance, $R_n$.\\
\indent Our squelch circuit has a Josephson junction between the signal wire and ground, mounted in the middle of a metal powder filter [Fig. \ref{fig1}(a),(b)]. The filter properties are measured by connecting it between a source and a load, i.e. $R_0=R_L=50$ $\Omega$ [Fig. \ref{fig1}(d)]. In a realistic application the load is determined by the following circuit element, e.g. a SQUID attached to a flux qubit \cite{squbits}. For small signals and for frequencies $f<f_{co}$ the transmission through the filter is given by the ratio $|T|=\omega L_J/\sqrt{\omega^2L_J^2+R_0^2}$ ; with $\omega L_J\ll R_0$ we obtain $T\propto\omega$, and $T\ll1$. For $f>f_{co}$ $T$ is governed by the metal powder properties [see Fig. \ref{fig1}(e)]. For larger signals the circuit behaves as a regular metal powder filter. Note that in this regime the Josephson junction carries a nonzero voltage and so it generates Josephson oscillations at the frequency $f_J=(2e/h)V$. By mounting the junction midway of the filter, this radiation is absorbed in the metal powder since $f_J\gg f_{co}$.\\ 
\indent Two different Josephson junctions made of aluminum have been used, with critical currents of 2.9 $\mu$A and 1 $\mu$A, respectively. Taking the junction with critical current $I_C=2.9$ $\mu$A, yields $L_J=110$ pH and $R_n=110$ $\Omega$. Experimentally we measure a copper powder filter cutoff frequency of 80 MHz (see below); for this frequency one calculates $A\approx0.001$ or 60dB for the small-signal attenuation by the junction inductance. The overall small-signal attenuation is smallest at ~200 MHz, with $A\approx50$dB, shown as the maximum in Fig. \ref{fig1}(e). Beyond $f_{co}$ we measure that the filter attenuation increases by approximately 65dB/decade.\\
\indent The metal powder filter is built by using a 2-component epoxy \cite{glue} loaded with copper powder. A thin ($\sim$0.1 mm) copper wire of 80 cm is meandered with a fixed spacing inside the epoxy-powder mixture, so that the impedance is uniform along the filter wire. The Josephson junction is made of a 40 nm film of aluminum, fabricated using electron beam lithography and double angle shadow mask evaporation [Fig. \ref{fig1}(c)]. The junction sits on a 400 $\mu$m thick thermally oxidized silicon substrate and is placed on a printed circuit board (PCB) in a gap left in the center of the copper powder filter. The PCB is attached to the housing of the copper powder filter with bolts to allow for good thermal anchoring and electrical grounding. Then it is soldered to the pins that connect to the copper wire, and the chip with the Josephson junction is glued on it with silver paint and wire-bonded (see Fig. \ref{fig1}a). Finally a metallic lid is tightly screwed onto the filter housing, preventing signal leakage.\\
\indent The $S$-parameters of the Josephson squelch were measured with a HP-8753C Network Analyzer at room temperature and at 4 K. In the transmission $S_{21}$ the \mbox{-3dB} point was seen at 80 MHz at room temperature, increasing to 100 MHz at 4 K. Beyond this frequency the transmission dropped steadily down by 50dB at 600 MHz. The reflection $S_{11}$ was seen to be -20dB up to 2 GHz, corresponding to an impedance of 60 $\Omega$.\\
\indent To characterize the performance of the Josephson squelch, we mounted it on the mixing chamber of a cryogen-free dilution refrigerator with a base temperature of 15 mK. We used a cryogenic high electron mobility transistor (HEMT) amplifier \cite{ivo} on the \mbox{0.6 K} still-stage and a post-amplifier at room temperature [Fig. \ref{fig2}(a)]. The post amplifier had a bandwidth of 5 kHz - 50 MHz. We employed optical isolation in the input and output of a HP-4195A Network-Spectrum analyzer together with a battery powered current source, which were combined in a bias-tee at the entrance of the dilution refrigerator.\\
\indent We first used the Josephson junction to probe the attenuation of the copper powder at dif\mbox{}ferent frequencies. We ramped a current $I_{DC}$ through the Josephson junction to monitor its IV characteristic [Fig. \ref{fig2}(b)]. For this particular junction we observed a critical current of 2.9 $\mu$A. On top of the ramping current we added a high frequency signal $I_{AC}$ of small amplitude, $I_{\mathrm{tot}}=I_{DC}+I_{AC}$. The AC signal reaching the junction experienced the attenuation of half the copper powder, which is frequency dependent. Signals of frequency $f\gg 2f_{co}$, with $f_{co}\simeq100$ MHz being the cutoff of the copper powder filter, were strongly damped and did not affect the IV characteristic (green curve in Fig. \ref{fig2}(b)). As the frequency was decreased below $2f_{co}$, the signals reaching the Josephson junction had increasing amplitude that was added to the slow current ramp, so that less $I_{DC}$ was needed to reach $I_C$. As a result, the critical current of the Josephson junction appeared to ef\mbox{}fectively decrease (as seen progressively in the brown, red and black traces in Fig. \ref{fig2}(b)).\\
\indent The squelch action was probed by applying a signal of small amplitude from the network analyzer (much smaller than $I_C$) and then adding a DC bias $I_{DC}$. In this measurement we used a Josephson junction with a critical current $I_C=1$ $\mu$A. As can be seen in Fig. \ref{fig3}, for $I_{DC}<0.8$ $\mu$A no signal beyond the noise floor was observed through the Josephson squelch. When the full current approached the critical current $I_{DC}+I_{AC}\approx I_C$, we measured a progressive increase on the voltage acquired by the amplifier. This is a direct proof of the performance of a Josephson junction used as a controllable squelch. When $I_{DC}\ge1.2$ $\mu$A, the Josephson junction was in its dissipative state and the signal saturated. In an actual quantum circuit application (see Fig \ref{fig1}(d)), when a load $R_L$ is employed behind the Josephson squelch, the pulse that probes the load needs to have a large enough amplitude so as to switch the junction to the voltage-carrying state.\\
\indent When the applied power was low enough, only the noise coming from the amplifier was observed. Taking the difference between the traces at 0 $\mu$A and at 1.2 $\mu$A in Fig. \ref{fig3}, the ON/OFF ratio was seen to be at least 50dB up to 50 MHz, where a resonance in some part of the circuit was observed.\\
\indent In summary, we have developed a squelch circuit that integrates a high frequency low-pass copper powder filter with a Josephson junction that acts as the signal-level-dependent element. The squelch operates from DC to more than 50 MHz with an ON/OFF ratio of more than 50dB. The Josephson squelch is highly attractive for the operation of quantum nanocircuits, where the interaction between instrumentation and nanocircuit needs to be active only during specific short periods while optimal isolation is required at all other times. In particular dephasing due to low frequency noise in control and measurement lines can be strongly suppressed.\\
\indent This work was supported by the Dutch Organization for Fundamental Research on Matter (FOM), by the EU project \mbox{EuroSQIP} with the Grant No. IST-3-FP6-015708-IP and by the NanoNED program.

\bibliographystyle{plain}

\end{document}